\documentclass[prx,aps,superscriptaddress,showkeys,showpacs,twocolumn]{revtex4}
\usepackage{graphicx} 
\usepackage[usenames]{color}
\usepackage{amsmath,amssymb}
\usepackage{gensymb}
\usepackage{natbib}
\usepackage{xcolor}
\def\strutdepth{\dp\strutbox}
\def\nw#1{\strut\vadjust{\kern-\strutdepth\vtop to0pt{\vss\hbox to\hsize
{\hskip\hsize\hskip5pt$\leftarrow$\hss\strut}}}{\em #1}}
\usepackage{rotating}

\pdfoutput=1

\begin{document}

\title{Rheology of granular flows across the transition from soft to rigid particles}
\author{Adeline Favier de Coulomb}
\affiliation{Physique et M\'ecanique des Milieux H\'et\'erog\`enes, PMMH UMR 7636 ESPCI -- CNRS -- Univ.~Paris-Diderot -- Univ.~P.M.~Curie, 10 rue Vauquelin, 75005 Paris, France}
\author{Mehdi Bouzid}
\affiliation{Department  of  Physics,  Institute  for Soft  Matter  Synthesis and Metrology, Georgetown  University,  37th and O Streets,  N.W., Washington,  D.C. 20057,  USA}
\author{Philippe Claudin}
\affiliation{Physique et M\'ecanique des Milieux H\'et\'erog\`enes, PMMH UMR 7636 ESPCI -- CNRS -- Univ.~Paris-Diderot -- Univ.~P.M.~Curie, 10 rue Vauquelin, 75005 Paris, France}
\author{Eric Cl\'ement}
\affiliation{Physique et M\'ecanique des Milieux H\'et\'erog\`enes, PMMH UMR 7636 ESPCI -- CNRS -- Univ.~Paris-Diderot -- Univ.~P.M.~Curie, 10 rue Vauquelin, 75005 Paris, France}
\author{Bruno Andreotti}
\affiliation{Physique et M\'ecanique des Milieux H\'et\'erog\`enes, PMMH UMR 7636 ESPCI -- CNRS -- Univ.~Paris-Diderot -- Univ.~P.M.~Curie, 10 rue Vauquelin, 75005 Paris, France}

\begin{abstract}
The rheology of dense granular flows is often seen as dependent on the nature of the energy landscape defining the modes of energy relaxation under shear. We investigate numerically the transition from soft to rigid particles, varying $S$, their stiffness compared to the confining pressure over three decades and the inertial number $I$ of the shear flow over five decades. We show that the rheological constitutive relation, characterized by a dynamical friction coefficient of the form  $\mu(I)=\mu_c + a I^{\alpha}$ is marginally affected by the particle stiffness, with constitutive parameters being essentially dependent on the inter-particle friction. Similarly, the distribution of local shear rate mostly depends on the inertial number $I$, which shows that the characteristic timescale of plastic events is primarily controlled by the confining pressure and is insensitive to $S$. By contrast, the form under which energy is stored between these events, and also the contact network properties such as the coordination number and the distance to isostaticity, are strongly affected by stiffness, allowing us to discuss the different regimes in the $(S,I)$ phase space.
\end{abstract}

\pacs{47.57.Gc, 83.80.F}



\date{\today}

\maketitle

\section{Introduction}
Soft amorphous materials are disordered assemblies of interacting particles that can resist shear like a solid, but flow like a liquid under a sufficiently large applied shear stress \cite{CPB09}. Controlling and tuning this solid-liquid transition is fundamentally important in many industrial processes involving paste, suspensions or emulsions. From a fundamental perspective, this dynamical transition was identified as a critical phenomenon associated with diverging length scales quantifying the degree of cooperativity of particle motion in the flow \cite{HW12,DJvH}. In particular, many experiments on granular materials and emulsions have shown that the rheology is actually non-local, i.e. the stress at a given location depends not only on the local strain rate but also on the mobility in a surrounding region (see \cite{Bouzid15b} and references therein).

The solid-like state is usually associated with highly multi-stable energy landscapes. For foams and emulsions, the free energy derives from surface tension while for soft particles  \cite{CBML03} or metallic glasses it results from the elasticity of the material \cite{CBML03}. Colloidal suspensions, with their entropic landscape, may belong to the same class of systems \cite{HW12,Ikeda12}. Elasto-plastic theories assume that the material behaves most of the time like a solid, but presents local and short lived elasto-plastic events~\cite{TLB06,LC09,BCA09,LP09}, in regions sometimes called 'shear transformation zones' \cite{FL98} or Eshelby flips. The prevailing picture is that when the material is sheared in a quasi-static manner, at vanishing shear rate $\dot \gamma$, the energy is slowly stored and rapidly released through scale-free avalanches \cite{DBZU11,Liu2016}, in close analogy with the depinning transition of an elastic line.

\begin{figure}[t!]
\includegraphics{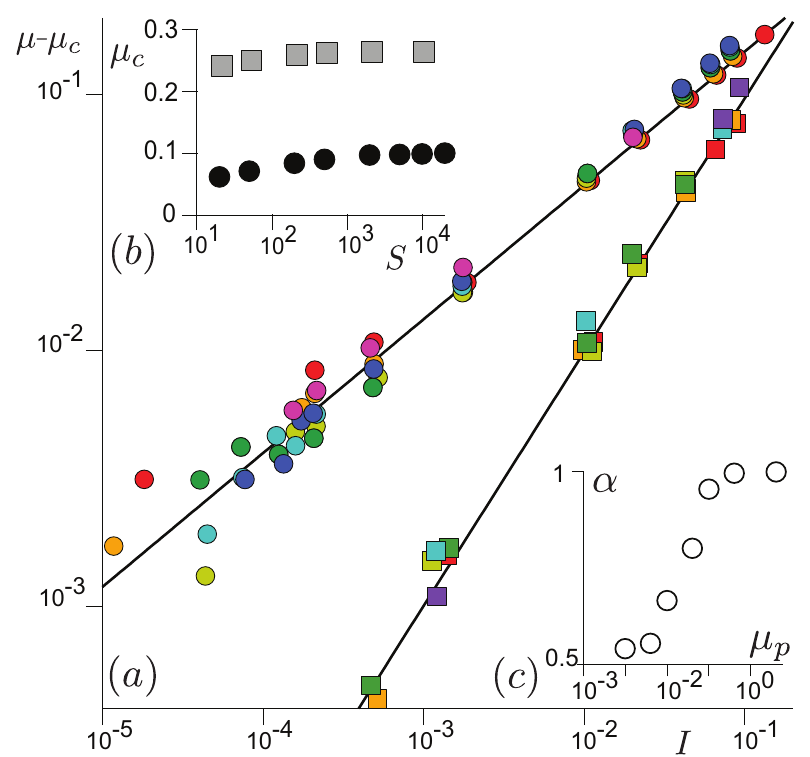}
\vspace{- 4 mm}
\caption{(a) Reduced macroscopic friction $\mu -\mu _c$ as a function of $I$. Circles: frictionless case. Squares: frictional case ($\mu _p=0.4$). Colors represent the value of $S$ going from $20$ (red) to $2\cdot10^4$ (pink), see the color code in Fig.~\ref{Fig4}. Solid lines: power laws with exponents $0.5$ and $1$, respectively for frictionless and frictional particles. (b) Variation of $\mu _c$ with $S$ in both cases (same symbols). (c) Variation of the exponent $\alpha$ with the microscopic friction coefficient $\mu _p$.}
\vspace{- 5 mm}
\label{Fig1}
\end{figure}

\begin{figure}[t!]
\includegraphics{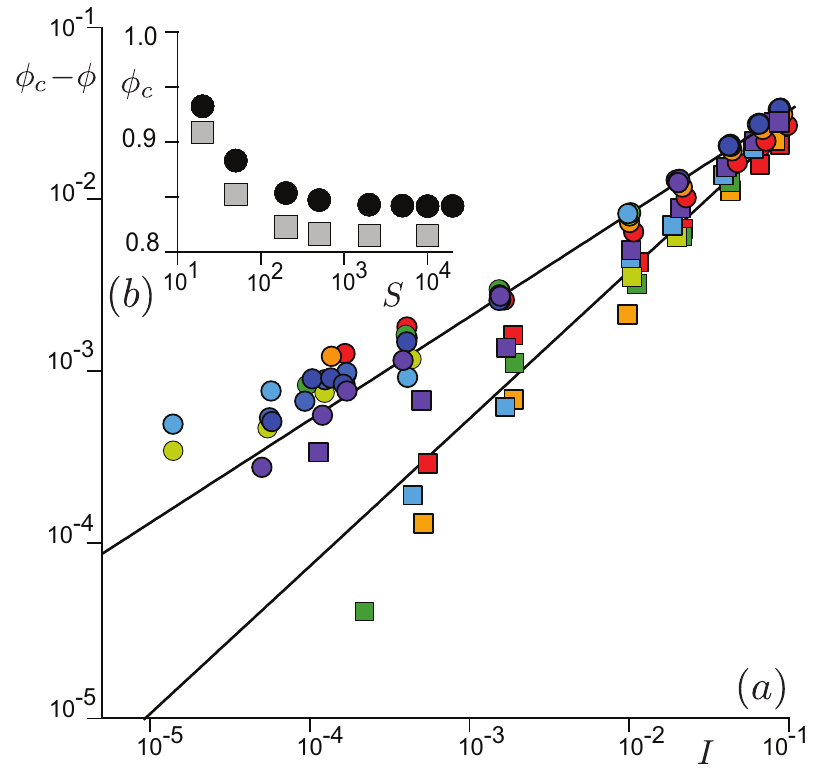}
\vspace{- 4 mm}
\caption{(a) Reduced packing fraction $\phi _c-\phi$ as a function of $I$. Circles: frictionless case. Squares: frictional case ($\mu _p=0.4$). The stiffness $S$ is represented by the same color code as in Fig.~\ref{Fig1}. Solid lines: power laws with exponents $0.6$ and $0.8$, respectively for frictionless and frictional particles. (b) Variation of $\phi _c$ with $S$ in both cases (same symbol convention).}
\vspace{- 5 mm}
\label{Fig2}
\end{figure}

By contrast, the energy landscape for an assembly of rigid grains is purely geometrical \cite{R00,Olsson07,LDW12}. The jamming approach, developed for hard non-deformable grains in a quasi-static state ($\dot \gamma \to 0$), is based on the topological analysis of the contact network. The control parameter of the rigidity transition is the mean number of contacts per grain $Z$ (see \cite{MVH} and references therein). Close to the jamming transition, flow is only possible along soft floppy modes, by essence spread in space and prescribing a cooperative motion of the particles \cite{LDW12,ABH12}.

These two separate quasi-static theories raise three questions whose responses constitute the core of this paper. (i) What is the effect of inertia? (ii) What does the rheology retain of the transition from soft to rigid particles? (iii) Is there any signature of the transition between the behavior of soft and hard particles? We address them here by means of numerical simulations.

\section{Numerical set-up}
The numerical system is a two-dimensional shear flow, similar to that used in~\cite{BTCCA13}, constituting of  $N = 2 \cdot 10^3$ grains whose diameter is picked up at random between $0.8 d$ and $1.2 d$ ($d$ is the average grain size) to avoid both segregation and crystallization. The dynamical equations are integrated using the Verlet algorithm. The contact between particles is standardly modeled by viscoelastic pistons, with a Coulomb friction along the tangential direction \cite{Luding06}. The normal spring constant $k_n$ is the first key parameter of the study. The tangential spring constant is $k_t=0.5 k_n$ and the Coulomb friction coefficient is varied from $\mu_p=0$ (frictionless case) to $\mu_p=4$. The value $\mu_p=0.4$ is used to illustrate the frictional case in figures. Damping parameters are chosen such that the restitution coefficient is $e \simeq 0.5$. We have checked that our results are independent of the value of $e$ between $e=0.1$ and $e=0.9$.

The flow is confined between two rough solid walls distant on the average by $H \simeq 50 d$. The top wall is moving along the $x$-direction at constant velocity whereas the bottom wall is motionless. Periodic boundary conditions are used along the $x$-direction, and we denote by $W$ the cell's width. The vertical position of the top wall is adjusted at each time-step to keep the confining pressure $P$ constant. This normal stress -- not to be confused with the trace of the stress tensor, although quantitatively very close (within a few percent) -- is indeed the relevant rescaling quantity for the granular rheology \cite{GDRMidi}. Simulations for which the vertical position of both walls are fixed are equivalent in the limit of large systems. In practice, however, for systems on the order of a few thousand grains, working at constant volume does not allow us to study very low shear rates, as jamming events generate huge fluctuations in this limit. This is one of the reasons for which the simulations described in \cite{GC16,C05}, which have investigated several aspects of the role of particle stiffness, are limited to the high shear rate regime.

We have paid a great deal of attention to the issue of statistical convergence of the quantities measured. Here we focus on the steady state, where the normal and shear stresses $P$ and $\tau$, and thus the friction coefficient $\mu=\tau/P$, are homogeneous across the cell. We thus ignore the effect of initial preparation and of hysteresis, where localized plastic events play a role in the aging properties and the shear-band localization dynamics \cite{BAU11,ANBCC12,LeBouil14}. As the double limit of vanishing shear rate and rigid particles turns out to be a critical point, slow relaxations are encountered and reaching a true steady state is a rather demanding computer task. In practice, the present results typically represent one year of simulations on one computer core. We have carefully checked that convergence was reached for all data points, which then corresponds to rather large values of the total strain $\dot\gamma t$.

\begin{figure}[t!]
\includegraphics{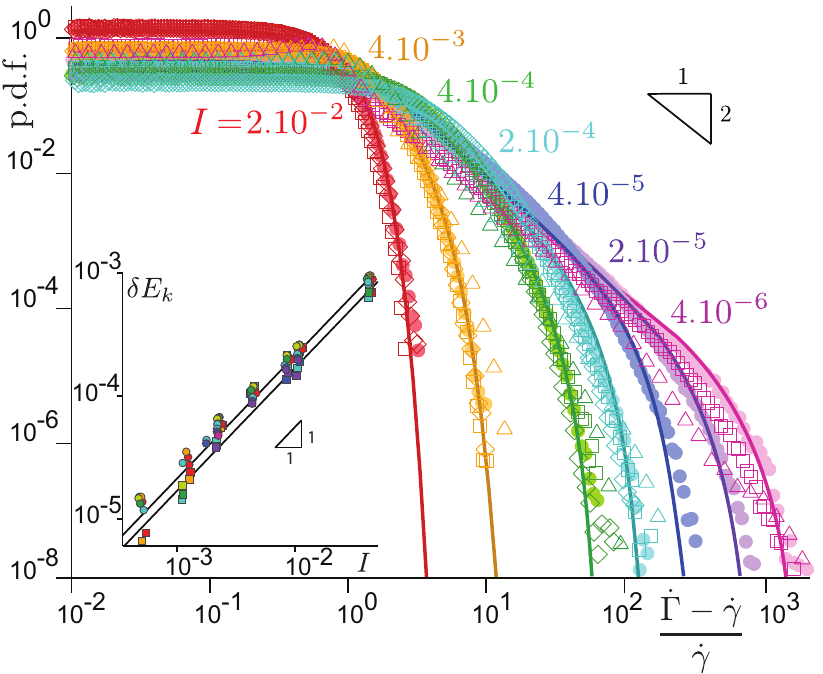}
\vspace{- 4 mm}
\caption{Probability distribution function (p.d.f) of the rescaled local shear rate $|\dot\Gamma-\dot\gamma|/\dot\gamma$ for different values of $I$ and $S$ (frictionless case: squares $S=50$, circles $S=200$, diamonds $S=2000$; frictional case: triangles $S=200$). Solid lines: phenomenological fit (Appendix.~\ref{OSM}). Inset: average fluctuating kinetic energy per grain $\delta E_k$, for a system composed of frictionless (circles) or frictional (squares) grains, as a function of $I$. The color code for $S$ is the same as in Fig.~\ref{Fig4}.}
\vspace{- 5 mm}
\label{Fig3}
\end{figure}

\section{Rheology across the transition from soft to rigid particles}
In the limit of a large number of particles, the system is characterized by two dimensionless numbers: the dimensionless stiffness $S=k_n/P$, which controls the inter-penetration of grains in contact and the inertial number $I=\dot \gamma \sqrt{m/P}$ which compares the shear rate $\dot \gamma$ to the plastic rearrangement time-scale $\mathcal{T}= \sqrt{m/P}$, where $m$ is the average mass of the grains. The rheology of the granular flows is here systematically investigated as a function of $S$ and $\mu_p$. Provided they are measured at long times, achieving statistical convergence, all data follow the standard constitutive relation \cite{GDRMidi,CEPRC05,JFP06}, which is well fitted by $\mu(I) = \mu_c + a I^\alpha$ (Fig.~\ref{Fig1}). We systematically checked that the rheology was monotonic at small $I$. The asymptotic value $\mu_c$ in the quasi-static limit is larger for larger $\mu_p$ \cite{dC04,KK14}. It also gently increases at small $S$, and saturates above $S \simeq 10^3$ (Fig.~\ref{Fig1}b). Except for this weak variation of $\mu_c$, one does not observe any signature of the transitions identified before on the rheology: Fig.~\ref{Fig1} shows that, for a given value of $\mu_p$, all reduced data points $\mu(I)-\mu_c$ collapse on a master curve. The only parameter having a strong influence is in fact the particle friction coefficient $\mu_p$: the exponent $\alpha$ sharply changes around $\mu_p=0.04$ from $0.5$ in the frictionless limit ($\mu _p=0$) to $1$ in the frictional case (Fig.~\ref{Fig1}c, see also (Appendix.~\ref{OSM})). This can be interpreted as the crossover when dissipation by frictional sliding becomes dominant in front of dissipation by collisions \cite{DMW16}.

Similarly, Fig.~\ref{Fig2} shows the reduced packing fraction $\phi _c -\phi$ as a function of the inertial number $I$ for both frictionless and frictional ($\mu _p=0.4$) cases. $\phi _c -\phi$ is also independent of $S$ and follows a power law dependence on $I$. However, as one can expect, the asymptotic value $\phi_c$ significantly depends on the grain elasticity, as illustrated by the inset in Fig.~\ref{Fig2}: it decreases with $S$ and reaches a plateau above $S \simeq 10^3$. This shows again the challenge of dealing with constant volume simulations in the rigid limit: at small shear rate, this requires $\phi$ to approach a finite value of $\phi_c$ with enough precision, simulating systems large enough not to stop due to critical fluctuations. By contrast, in a  controlled-pressure set-up, the critical state is naturally approached in the limit $I \to 0$.

The microscopic and mechanistic foundation of these constitutive relations, which would result from first principles, is still a challenging open issue. A major step has been to show that non-Brownian suspensions close to jamming are also described by a pressure-controlled rheology, totally analogous to that of inertial (dry) grains \cite{BGP11,TAC12}. Because these suspensions follow an overdamped dynamics, their rheology is directly related to non-affine collective motions \cite{ABH12}. These cooperative effects are purely geometrical and have been related to an anomalous mode of a spring network that would have an identical geometry \cite{LDW12}. Statistical properties of trajectories are essentially determined by steric effects and are insensitive to the details of dissipative mechanisms. Interestingly, the rheology of Brownian suspensions can also be mapped to that of non-Brownian suspensions of soft elastic spheres, replacing thermal noise by mean field entropic repulsions \cite{TBKCCA15,BW06}. The analogy between constitutive relations of grains in the overdamped and inertial regimes thus strongly suggests that steric effects control the statistical properties of grain trajectories in dense granular flows, and these results showing the quasi-independence of the law $\mu(I)$ with respect to the grain stiffness reinforce this idea.
 
\section{Local shear rate distribution}
To investigate the origin for the very weak dependence of the rheology $\mu(I)$ on the stiffness, we measure $\dot \Gamma (\vec{r},t)$, which evaluates the local rate of deformation at position $\vec{r}$ and time $t$. As in \cite{Bouzid15b}, we take for the coarse-graining function a Gaussian of width $\kappa$ and define $\dot \Gamma$ at the location $\vec {r}_i=(x_i,z_i)$ of the grain $i$ as 
\begin{eqnarray}
\dot\Gamma(\vec{r}_i,t)=\frac{\sum \limits_{j=1}^N(u_i-u_j)(z_i-z_j)\exp\left(-\frac{||\Delta\vec{r}||^2}{2\kappa^2}\right)}{\sum \limits_{j=1}^N(z_i-z_j)^2\exp\left(-\frac{||\Delta\vec{r}||^2}{2\kappa^2}\right)} \, ,
\label{Gamma}
\end{eqnarray}
where $u_i$ (resp. $u_j$) is the horizontal velocity of the grain $i$ (resp. grain $j$) at time $t$, and $||\Delta \vec{r}||=\sqrt{(z_i-z_j)^2+(x_i-x_j)^2}$ defines the distance between the grains $i$ and $j$ at time $t$. In all the results presented in this paper, we have set $\kappa = d$.

By construction, its time average is everywhere equal to $\dot \gamma$, but the spatial and temporal values of $\dot \Gamma$ can be largely distributed. We display in Fig.~\ref{Fig3} the probability distribution function of $\dot \Gamma$, rescaled by $\dot \gamma$, for different values of $I$ and $S$. We observe that the distributions are mostly independent of the stiffness $S$ and are controlled by the inertial number $I$, in both the frictionless and the frictional cases. This is consistent with the idea that granular trajectories are mostly determined by steric effects \cite{ABH12}. At large $I$, the distribution is a narrow Gaussian, reflecting permanent fluctuations around the average linear shear flow. Conversely, for asymptotically small $I$ the distribution shows an algebraic tail decreasing as $\dot \Gamma^{-2}$. This scale-free behavior is reminiscent of the intermittent dynamics associated with scale-free avalanches of plastic events \cite{LC09,LP09,DBZU11,Maloney15}. Here, however, the same behavior is recovered in the rigid limit, i.e. close to jamming, where the non-affine motions takes place along soft modes. The power law distribution of $\dot \Gamma$ presents an inner cut-off at the time-scale $\dot \gamma^{-1}$, which is the average time between plastic events. The large shear rate cut-off, which corresponds to the fastest granular reorganizations, is determined by the time-scale $\mathcal T$. This means that the confining pressure provides the relevant energy scale for plastic events whatever the stiffness. The crossover between the two regimes can be defined as the point at which the algebraic tail disappears, i.e. when the standard deviation is on the order of the mean: $\delta \dot\Gamma = \dot\gamma$. As evidenced in Fig.~\ref{Fig3}, this corresponds to a value of $I$ around $2 \cdot 10^{-3}$, independently of $S$.

These measurements suggest a simple picture of plastic events, based neither on elasto-plasticity nor on jamming. The grains move with short phases of typical duration $\mathcal{T}$ during which they are suddenly accelerated by the pressure $P$ to a velocity $d/{\mathcal T}$ with respect to their neighbors. On the average, however, this relative velocity is equal to $\dot \gamma d$, which is much lower in the limit of small $I$. The average fluctuating kinetic energy per grain (related to the so-called granular temperature) then scales as $m (d/{\mathcal T})^2$, multiplied by the fraction of time during which this acceleration phase occurs, that is $\dot\gamma {\mathcal T} = I$ \cite{GDRMidi}. This argument therefore gives $\delta E_k \sim d^2 P I$, a prediction nicely obeyed by our measurements (inset in Fig.~\ref{Fig3}). By contrast, the energy stored in the springs representing the contacts is controlled by the typical overlap $d/S$ between grains, which gives a potential energy per grain scaling as $E_p \sim k_n (d/S)^2$. This ignores the dynamical pressure due to the collisions between the grains, which scales as $I^4$. We find that the fluctuating potential energy $\delta E_p$ is a fraction of $E_p$, following the same scaling in the limit $I \to 0$ (Appendix.~\ref{OSM}), so that the ratio $\delta E_k/\delta E_p$ scales as the product $SI$ at small $I$. This confirms that plastic events keep the same time and velocity scales across the transition from deformable to rigid grains. Potential energy is the dominant form of energy at small $SI$ and kinetic energy at large $SI$.

To sum up this section, the key ingredient to understanding the rheology turns out to be the fluctuations of the local shear rate i.e. the motion intermittency. It reveals the existence of scale free processes in time, insensitive to the particle softness. The long time cut-off is set by the average time $\dot \gamma^{-1}$ needed for grains to move by their own size. The short time cut-off $\mathcal T$ is set by the time needed to perform the plastic motion, which is controlled by the normal stress $P$ whatever the form of energy dominating the energy landscape.

\begin{figure}[t!]
\includegraphics{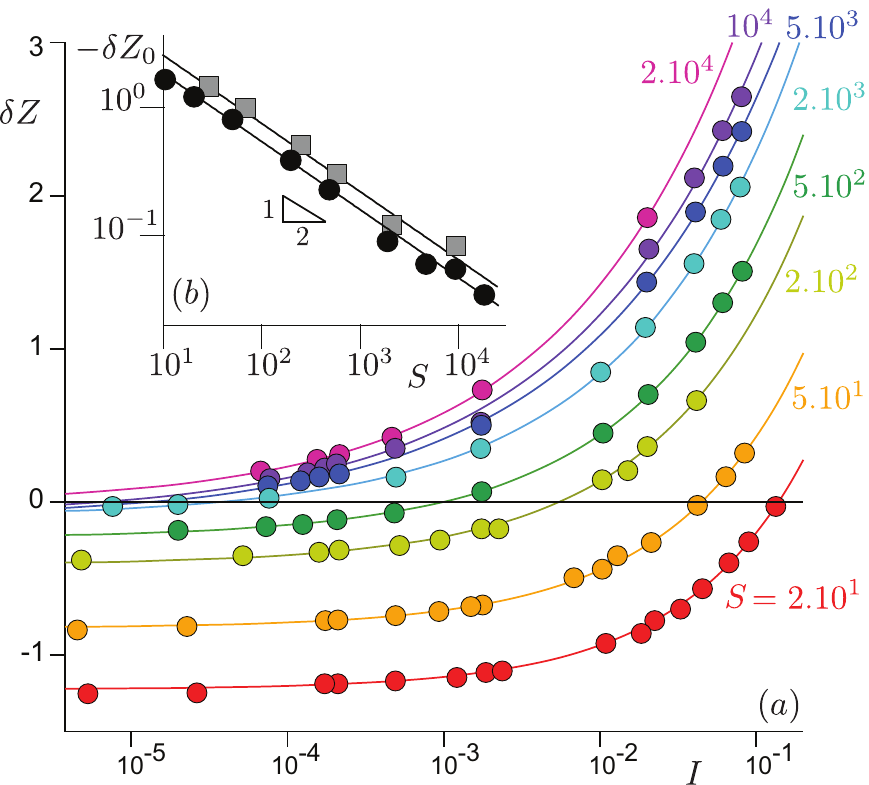}
\vspace{- 4 mm}
\caption{(a) Distance to isostaticity measured for frictionless grains for different values of $S$ (see legends). Solid lines: best fit by a (negative) asymptotic constant $\delta Z_0$ plus a power law of $I$. (b) Variation of $-\delta Z_0$ for the frictionless (circles) and frictional (squares) cases with $S$. Solid lines: power law fits as $S^{-1/2}$.}
\vspace{- 5 mm}
\label{Fig4}
\end{figure}

\section{Isostatic transition}
Having shown that both the rheology and the distribution of local shear rate are mostly insensitive to the particle stiffness, our aim is now to investigate other characteristics of the granular system, sensitive to $S$. The jamming theory is based on the Maxwell rigidity criterion. We accordingly define the distance $\delta Z$ to isostaticity as the difference between the average number of constraints and the average number of degrees of freedom (the number of force components) per grain. One must distinguish between the frictionless ($\mu_p=0$) and frictional ($\mu_p>0$) cases:
\begin{equation}
\delta Z = \left\{
\begin{array}{ll}
2D-Z		& \quad {\rm when} \quad \mu_p=0,
\\
D+1+\zeta-Z	& \quad {\rm when} \quad \mu_p>0,
\end{array}
\right.
\end{equation}
where $D$ is the space dimension (here $D=2$) and $\zeta$ the fraction of sliding contacts \cite{MVH}.  A system is `solid', i.e. rigid, when hyperstatic ($\delta Z<0$) and `liquid' when hypostatic ($\delta Z>0$). Fig.~\ref{Fig4}a shows $\delta Z$ as a function of $I$ for different values of $S$. In the limit $I \to 0$, $\delta Z$ tends to a negative constant $\delta Z_0$ and increases as a power law of $I$ whose exponent, close to $0.5$, increases slightly but systematically with $S$. The limit of vanishing shear rate, it varies as $-\delta Z_0 \sim S^{-1/2}$ for both frictional and frictionless grains (Fig.~\ref{Fig4}b). This scaling law originates from the pair distribution function at the jamming point. As the probability to observe a gap $\Delta$ between rigid particles at the jamming point diverges as $\Delta^{-1/2}$ \cite{MVH}, the creation of new contacts between jammed soft grains when compressed must scale as $S^{-1/2}$, in line with the result of \cite{SGBL07} for gravity-driven flows.

The set of curves $\delta Z(I)$ shows that, at finite stiffness $S$, the isostatic transition $\delta Z=0$ is crossed at a finite value of $I$. Comparing Figs.~\ref{Fig1} and \ref{Fig4} we then see that the rheology is insensitive to this transition from hyperstatic to hypostatic.  Interestingly, though, we show that the double limit of rigid grains $S\to\infty$ at vanishing shear rate $I\to0$ coincides with isostaticity for both frictionless and frictional particles. Such a packing in the critical state associated with an isostatic contact network is asymptotically reached under shearing during a time that diverges with the size of the sample.

\begin{figure}[t!]
\includegraphics{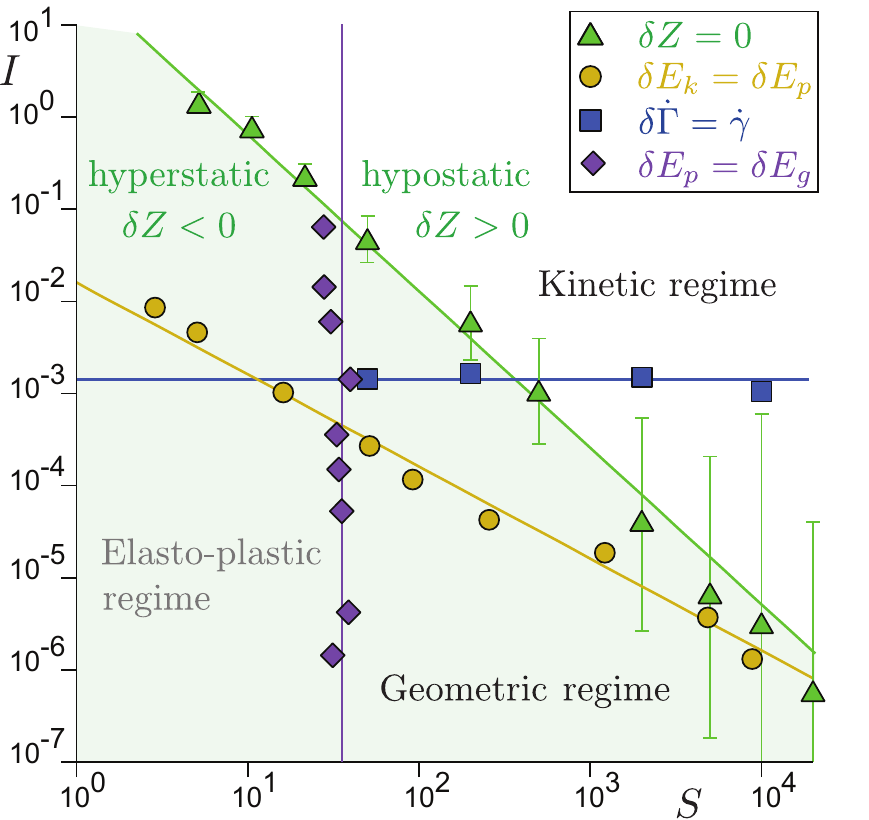}
\vspace{- 4 mm}
\caption{Transitions in the plane relating the inertial number $I$ to the stiffness $S$ in the frictionless case. The symbols show the location of the transition from the elastic to the kinetic regime (yellow circles), the transition from the elastic to the geometric regime (violet diamonds), the isostatic criterion (green triangles) and the presence of tails in the distribution of the local shear rate (blue squares). Solid lines show the best fits by power laws.}
\vspace{- 5 mm}
\label{Fig5}
\end{figure}

\section{Cross-overs between regimes}
It is instructive to summarize the different results of this paper in the $(S,I)$ plane, featuring cross-overs between the different regimes we have discussed above. A first line is the isostatic condition $\delta Z=0$. It is reported with green triangles in Fig.~\ref{Fig5}, and follows the empirical power law $I \propto S^{-\beta}$. It separates hypostatic contact networks (on the upper right) from hyperstatic ones (on the lower left). The exponent $\beta \simeq 1.7$ is found to be independent of $\mu_p$, within error bars (Appendix.~\ref{OSM}). The other lines of Fig.~\ref{Fig5} corresponds to the comparison of the forms under which energy is transiently stored between plastic events. The crossover from a potential energy dominated regime to a kinetic energy dominated regime occurs for $\delta E_k = \delta E_p$. According to the scaling laws discussed above, it scales as $I \propto S^{-1}$. The work of the confining pressure leads to an accumulation of energy of geometrical origin, by dilation of the system. It is characterized by a typical variation $\delta E_g = P W \delta H/N$, where $\delta H$ is the standard deviation of the cell's height. Consistently with a purely geometrical origin, $\delta H$ is found to be independent of both $S$ and $I$, and is on the order of a fraction of the grain size: $\delta H \simeq 0.1 d$ (Appendix.~\ref{OSM}). The crossover from a potential energy dominated regime to a geometric energy dominated regime occurs when $\delta E_p = \delta E_g$, which is, for this system size, around $S \simeq 30$ at small $I$ (Appendix.~\ref{OSM}). 

We can then conclude that elasto-plasticity takes place in the region $I \lesssim 10^{-3}$ where plastic events are separated from each others (presence of tails in Fig.~\ref{Fig3}), for particles soft enough (small $S$) to allow the potential energy variations $\delta E_p$ to dominate both kinetic $\delta E_k$ and geometric $\delta E_g$ energy modes. Consistently, this regime is in the hyperstatic domain ($\delta Z<0$). The relevance of the jamming theory is demonstrated in the double limit of vanishing $I$ and rigid particles (asymptotically large $S$), where isostaticity is recovered. At finite shear rate, however, typically above $I \gtrsim 10^{-3}$, all traces of criticality and elasto-plasticity disappear (vanishing power-law tails in Fig.~\ref{Fig3}): a permanently fluctuating regime appears, which does not exhibit any transition from soft to rigid particles.

The fact that the rheology is not sensitive to the particle stiffness, provided that it is expressed as $\mu(I)$, means that the pressure $P$ provides the relevant scale for the free energy density, independent of the energy landscape the grains are experiencing. These results thus extend the validity of the pressure-controlled rheology previously generalized from the inertial to the overdamped regime \cite{BGP11,TAC12,TBKCCA15}, or even to the case of `active' granular flows \cite{PCCA16}. In addition, we see that these crossovers are typically located in the lower part of the diagram of Fig.~\ref{Fig5}, hence the importance of the investigation of the low $I$ limit. This contrasts with the work of Campbell \cite{C05} which, using a volume-controlled set-up, has investigated the kinetic regime, typically $I \gtrsim 10^{-2}$, and has essentially identified the transition to the gaseous regime (Appendix.~\ref{OSM}).

\section{Concluding remarks on the origin of non-locality}
This work on the transition from soft to rigid particles was originally motivated by an apparently different problem: the origin of non-locality in dense granular materials, to which we will return to conclude this paper. So far, different phenomenologies have been proposed to take non-local effects into account in the rheology (see the review in \cite{Bouzid15b} and references therein). One starting point is to hypothesize that the friction coefficient $\tau/P$ depends not only on the local value of $I$ at the same location, but also on $I$ in the surrounding area. Then, by performing gradient expansion, one obtains the generic form of the constitutive relation, with a correction proportional to $\nabla^2 I/I$ \cite{BTCCA13}. An alternative starting point is to hypothesize that $I$ does not entirely describe the local state of the system and must be complemented by another state variable, the fluidity $f$. This parameter is generically governed by a reaction diffusion equation \`a la Ginzburg-Landau \cite{BCA09}, which essentially retain the symmetries of the problem. Although different expressions and interpretations of $f$ have been proposed ($f$ as the elastic strain, $f$ as the rate of plastic events, $f$ as a mechanical noise, etc), the very same linear equations at the perturbative order are obtained, which are insensitive to the dynamical mechanisms. In order to understand the foundation of the rheology of dense granular flows, one needs to go beyond the phenomenology and perform measurements that can shed some light on the mechanics of the problem.

The results presented here can be used to test two ideas proposed to explain non-locality of dense granular materials: elasto-plasticity on the one hand and non-affine modes controlled by the contact network on the other. Playing with two new control parameters (the particle softness and the contact friction), we have identified the region of parameter space $(S,I)$ in which elasto-plasticity is observed (Fig.~\ref{Fig5}). Because there are no elasto-plastic events in the rigid limit, the interpretation that non-locality is generated by such localized events interacting through the elastic stress or activated by a mechanical noise (force fluctuations \cite{PF09}, or velocity fluctuations \cite{ZK17}) cannot be correct for dense flows of hard grains. The fact that constitutive laws essentially ignore the transition from soft to hard grains rather supports the idea that the key ingredient to understand the rheology is motion intermittency \cite{SLJL10,RMME15}, as it reveals the existence of scale free processes in time, insensitive to the particle stiffness.

This suggests the exploration of further temporal fluctuations, and not only the geometry of trajectories as in the quasi-static theory of soft modes. Non-affine motions, even imperfectly related to the contact network properties, intrinsically provide a source of non-locality, as they capture the idea that the dissipation at a given location depends on the structure (the degree of fluidity) in the vicinity. Finally, we emphasize that inter-particle friction is the main ingredient leading to a substantial change in the rheology. This is related to hysteresis, whose understanding needs to focus on the relation between elasto-plasticity and jamming in systems prepared far from the steady sheared limit. A second perspective would be to bridge the gap with recent studies investigating the effect of inertia on the rheology of amorphous soft systems, in the limit $S \simeq 1$ \cite{KB16,KFB17}.

BA was supported by Institut Universitaire de France. This work was funded by a CNES grant.


\newpage
\appendix

\section{Supplementary Material}
\label{OSM}

We report below some supplementary figures and material, especially regarding frictional grains and data analysis.
 
\emph{Local strain rate $\dot\Gamma$} -- The probability distribution function of $|\dot \Gamma - \dot \gamma |/\dot\gamma$, which is displayed in Fig.~\ref{Fig3}, is fitted by the following function:
\begin{equation}
f(x)= A \frac{e^{-Bx^2}}{\left( 1+Cx^{a /b}\right) ^{b}} \, , 
\label{fit}
\end{equation}
with $x=\ln (|\dot \Gamma - \dot \gamma |/\dot\gamma)$, and where $A$, $B$, $C$, $a$, and $b$ are adjustable parameters. This phenomenological expression allows one to go continuously from a pure Gaussian to a Gaussian with algebraic tails.

\begin{figure}[h!]
\includegraphics{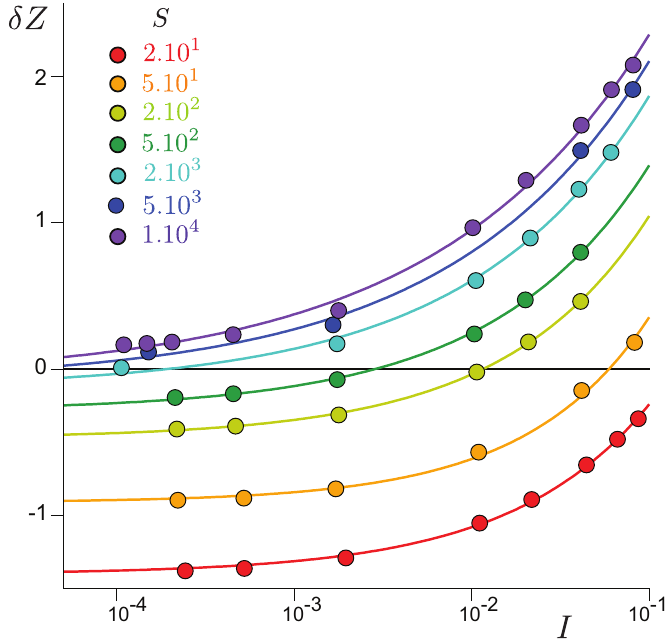}
\vspace{- 4 mm}
\caption{Distance to isostaticity measured for frictional grains. Colors represent different values of the stiffness $S$, ranging from $20$ (red) to $2 \cdot 10^{-4}$ (magenta), see legend. Solid lines: best fit by an asymptotic constant $\delta Z_0$ plus a power law of $I$. Black solid line: isostatic point $\delta Z=0$.}
\label{FigSI1}
\end{figure}

\begin{figure}[t!]
\includegraphics{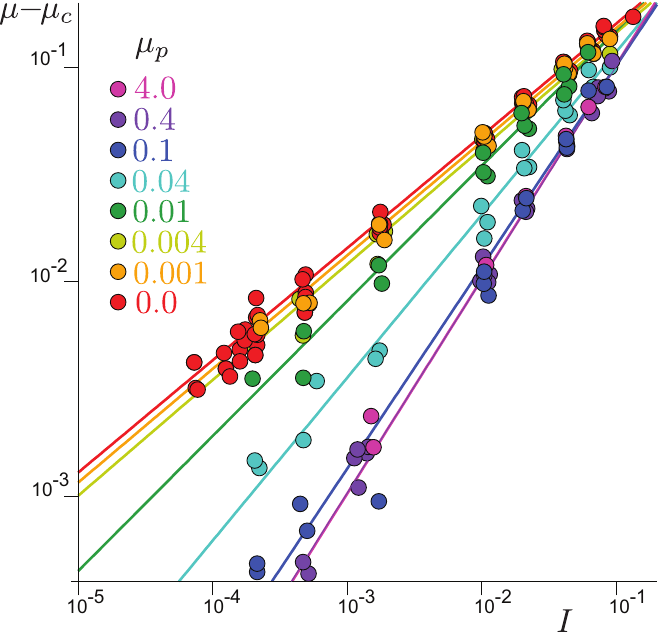}
\vspace{- 4 mm}
\caption{Reduced macroscopic friction $\mu -\mu _c$ as a function of $I$. The color encodes the microscopic friction $\mu_p$. Data from runs with different values of the stiffness are displayed and collapse. Solid lines: power laws with exponents varying from $0.5$ to $1$ (see Fig.~1c of the main paper).}
\label{FigSI2}
\end{figure}

\emph{Coordination number for frictional grains} -- Fig.~\ref{FigSI1} shows the variation of the distance to isostaticity $\delta Z$ with respect to $I$ in the frictional case. This is the equivalent of Fig.~\ref{Fig4}, which is for frictionless grains. The exponent of the law $\delta Z(I)$ is close to $0.7$, i.e. a bit larger than its value in the frictionless case, with a slight increasing trend with $S$.

\begin{figure}[t!]
\includegraphics{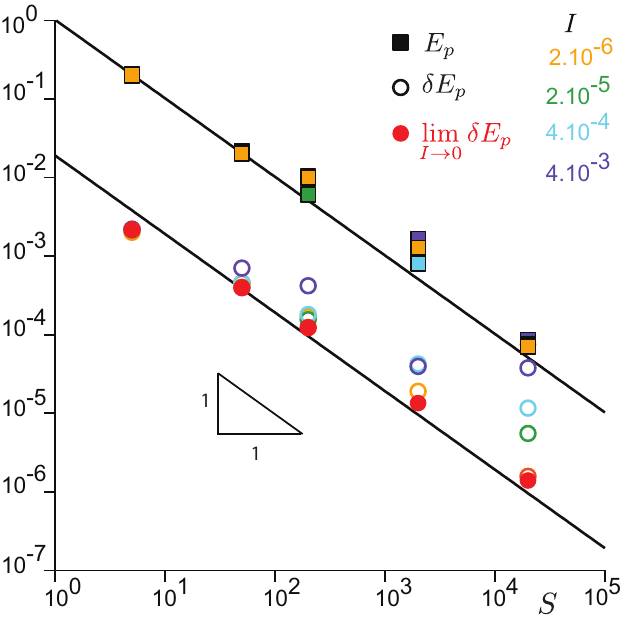}
\vspace{- 4 mm}
\caption{Potential energy per grain $E_p$ (squares) and its standard deviation $\delta E_p$ (circles), as a function of the stiffness $S$, and for different values of the inertial number $I$ (see color code in legend).}
\label{FigSI5}
\end{figure}

\begin{figure}[b!]
\includegraphics{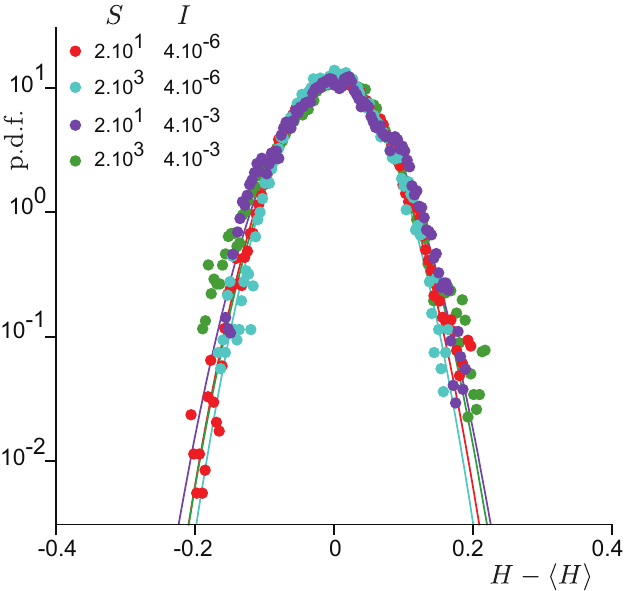}
\vspace{- 4 mm}
\caption{Probability distribution function (p.d.f) of the fluctuations of the system height $H-\langle H \rangle$, for different values of $I$ and $S$ (see legend). Solid lines: best gaussian fit.}
\label{FigSI3}
\end{figure}

\begin{figure}[b!]
\includegraphics{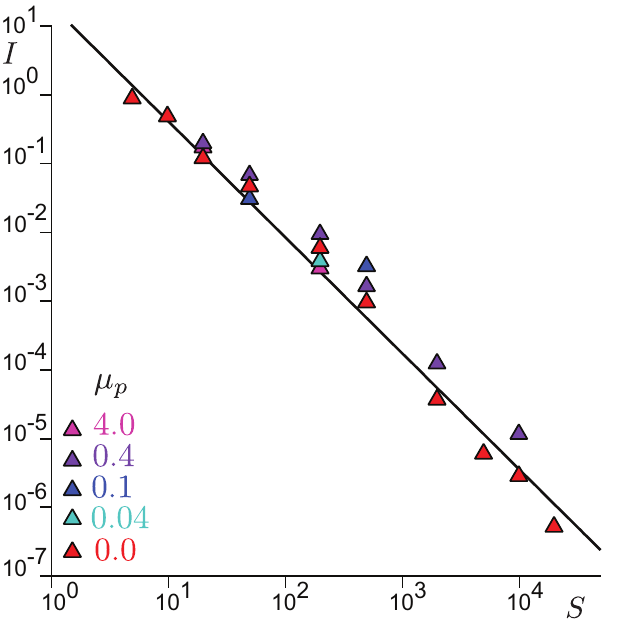}
\vspace{- 4 mm}
\caption{Isostatic criterion $\delta Z=0$ in the plane relating the inertial number $I$ to the stiffness $S$ for different values of the microscopic friction $\mu_p$ (see legend). Solid line: best fit by a power law $I \sim S^{-1.7}$.}
\label{FigSI4}
\end{figure}

\emph{Rheology for different grain frictions} -- Fig.~ \ref{FigSI2} shows the reduced macroscopic friction $\mu -\mu _c$ as a function of the inertial number $I$ for $8$ different values of the microscopic friction $\mu_p$ between $0$ and $4$. The data are independent of $S$ and scale with $I$ following a power law whose exponent increases with $\mu_p$, as displayed in Fig.~\ref{Fig1}c.

\emph{Potential energy} -- Fig.~\ref{FigSI5} shows the averaged potential energy $E_p$ per grain and the amplitude of its fluctuations $\delta E_p$ as a function of the stiffness $S$ for different values of the inertial number $I$. Both are found to scale as $S^{-1}$, at least in the limit of small $I$.

\emph{System height fluctuations} -- Fig.~\ref{FigSI3} shows the probability distribution function (p.d.f) of the fluctuations of the system height $H-\langle H \rangle$, for different values of $I$ and $S$. The amplitude $\delta H$ of these fluctuations, on the order of $0.1d$, is found to be independent of $I$ and $S$.

\emph{Isostaticity} -- Fig.~\ref{FigSI4} shows the isostatic criterion $\delta Z=0$ for different values of the microscopic friction $\mu _p$. This criterion is found to be independent of $\mu _p$.

\begin{figure}[t!]
\includegraphics{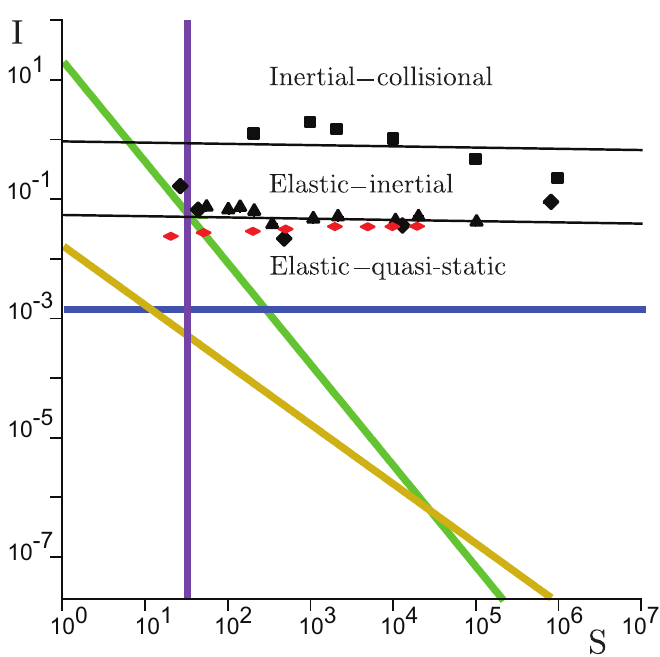}
\vspace{- 4 mm}
\caption{Transitions in the plane relating the inertial number $I$ to the stiffness $S$. Schematized from Fig.~\ref{Fig5}, the yellow solid line shows the location of the transition from the elastic to the kinetic regime, the violet one shows the transition from elastic to the geometric regime, the green one shows the isostatic criterion and the blue one shows the presence of tails in the distribution of the local shear rate. Black symbols represent Campbell's data interpreted by this author as the transitions between different regimes: elastic-quasistatic, elastic-inertial and inertial-collisional [21]. Red diamonds correspond to the values of $I$ for which $\mu = 2\mu_c$ for the same value of $S$, as determined from the simulations presented here.}
\label{FigSICampbell}
\end{figure}

\emph{Comparison with Campbell's transitions} -- Fig.~\ref{FigSICampbell} shows the two transitions identified by Campbell [21] in the plane $(S,I)$ presented in Fig.~\ref{Fig5}. According to this author, they correspond to limits between three regimes : elastic-quasistatic, elastic-inertial and inertial-collisional. The upper transition corresponds to what we would call the transition from dense to gaseous regime. It is mostly independent of $S$, as expected. The second transition, although claimed to depend on elasticity, is also found to be at a constant value of $I \simeq {\rm 8.10^{-2}}$. This line rather corresponds to the typical value of $I$ for which $\mu$ doubles with respect to $\mu_c$. Note that most of Campbell's numerical data, from which these transitions are deduced, have been obtained in the upper part of our diagram, i.e. in the region corresponding to the larger values of the inertial number ($I \gtrsim 10^{-2}$), outside the range of parameters where actual transitions take place.

\end{document}